\newcommand \bs{\begin{subequations}}
\newcommand \es{\end{subequations}}
\newcommand \bea{\begin{eqnarray}}
\newcommand \eea{\end{eqnarray}}
\newcommand \be{\begin{equation}}
\newcommand \ee{\end{equation}}
\newcommand \nn{\nonumber}

\documentclass[
 aps,
 prl,
 amssymb,
 twocolumn
 ]{revtex4}

\usepackage{graphicx}
\usepackage{amsmath}

\begin{document}

\title{Quantum noise memory effect of multiple scattered light}

\author{P. Lodahl} \affiliation{Research Center COM, NanoDTU, Technical
University of Denmark, Building 345V, Dk-2800 Lyngby, Denmark.}
\email{pel@com.dtu.dk}

\begin{abstract}
We investigate frequency correlations in multiple scattered light
that are present in the quantum fluctuations. The memory effect
for quantum and classical noise is compared, and found to have
markedly different frequency scaling, which was confirmed in a
recent experiment. Furthermore, novel mesoscopic correlations are
predicted that depend on the photon statistics of the incoming
light.
\end{abstract}
\date{\today}

 \maketitle

Light propagating through a disordered distribution of strongly
scattering particles induces an apparently random intensity
pattern known as a volume speckle pattern. Despite the apparent
randomness, such intensity speckle patterns can posses strong
temporal and spatial correlations, giving rise to the so-called
memory effect \cite{Feng88}. The memory effect manifests itself as
strong correlation between different spatial directions or
frequency components of the speckle pattern. After its
demonstration in early experiments \cite{Freund88,Garcia91}, the
memory effect has been employed as a sensitive technique of
measuring the diffusion constant of light propagation
\cite{Vellekoop05}. For extremely strong scattering, close to the
transition to Anderson localization, classical intensity
correlations can be strongly enhanced, which are referred to as
mesoscopic correlations \cite{Scheffold98,Chabanov00}. Recently,
it was shown that speckle correlations also exist when recording
the fluctuations of light instead of merely the intensity
\cite{Lodahl05-PRL}. Furthermore, strong spatial quantum
correlations were predicted depending on the quantum state of
light incident on the medium \cite{Lodahl05-corr}. In the present
Letter we derive the noise correlation function, which was
measured in \cite{Lodahl05-PRL}, for an arbitrary quantum state of
light. Our work demonstrates that the quantum fluctuations of a
volume speckle pattern give rise to a strong memory effect that
behaves different than classical fluctuations and also different
than the classical memory effect for intensities.

We study the propagation of photons through a multiple scattering
random medium. Using the quantum model for multiple scattering
\cite{Beenakker98}, we discretize the system into $N$ different
input modes and output modes. Let $\hat{n}_{\omega}^{ab}$ denote
the photon number operator for light at the frequency $\omega$
propagating from input channel $a$ to output channel $b$. The
fluctuations are quantified by the photon number variance $(\Delta
n_{\omega}^{ab})^2 = \left< ({\hat{n}_{\omega}^{ab}})^2 \right> -
\left< {\hat{n}_{\omega}^{ab}} \right>^2,$ where the quantum
mechanical expectation value will be evaluated for different
quantum states of light. Further details about this model are
given in \cite{Lodahl05-corr}. The quantum noise frequency
correlation function for light propagating from input channel $a$
to output channel $b$ is defined by
\be C_{ab}^N(\Delta \omega) = \frac{\overline{(\Delta
n_{\omega}^{ab})^2 \times (\Delta n_{\omega+\Delta
\omega}^{ab})^2} }{\overline{(\Delta n_{\omega}^{ab})^2}^2}-1. \ee
This function measures the correlation between the fluctuations in
a specific output direction for a frequency offset of $\Delta
\omega$, and is the generalization of intensity correlation
functions to the case of photon number fluctuations. The bars
above the variances denote averaging over ensembles of disorder.
We consider an arbitrary quantum state of light coupled through
channel $a$ characterized by an averaged number of photons
$\left<\hat{n}_{\omega}^a \right>$ and a Fano factor $F_a =
(\Delta n_{\omega}^{a})^2/\left<\hat{n}_{\omega}^a \right>$. The
photon number fluctuations of light transmitted to channel $b$ is
given by \cite{Lodahl05-corr}
\be \left[ {(\Delta n_{\omega}^{ab})^2} \right]_{QN} =
\left<\hat{n}_{\omega}^a \right> {T_{\omega}^{ab}} +
\left<\hat{n}_{\omega}^a \right> (F_a-1)
 {(T_{\omega}^{ab})^2}, \ee
where $T_{\omega}^{ab}$ is the intensity transmission coefficient
from channel $a$ to channel $b$. Here the subscript QN refers to
quantum noise of the relevant quantum state. In the situation
where classical noise (CN) is dominating, which, e.g., is the case
when modulating a laser beam, the variance of the fluctuations is
proportional to the squared average number of photons
\cite{Lodahl05-PRL}
\be \left[ {(\Delta n_{\omega}^{ab})^2} \right]_{CN} \propto
\left<\hat{n}_{\omega}^a \right>^2 {(T_{\omega}^{ab})^2}. \ee
Consequently, the correlation functions for quantum and classical
noise are given by
\begin{widetext}
\bs \bea C_{ab}^{QN}(\Delta \omega) &=& \frac{\overline{T_{\omega}
T_{\omega+\Delta \omega}} + (F_a-1) \left[ \overline{T_{\omega}^2
T_{\omega+\Delta \omega}} + \overline{T_{\omega} T_{\omega+\Delta
\omega}^2} \right] + (F_a-1)^2 \overline{T_{\omega}^2
T_{\omega+\Delta \omega}^2} }{\overline{T_{\omega}}^2 + 2 (F_a-1)
\overline{T_{\omega}} \times \overline{T_{\omega}^2} + (F_a-1)^2
\overline{T_{\omega}^2}^2  } -1,
\label{Corr-fcts-a} \\
C_{ab}^{CN}(\Delta \omega) &=& \frac{\overline{T_{\omega}^2
T_{\omega+\Delta \omega}^2} }{\overline{T_{\omega}^2}^2}-1,
\label{Corr-fcts-b}\eea \es
\end{widetext}
where we have assumed $\left< \hat{n}_{\omega}\right> = \left<
\hat{n}_{\omega+\Delta \omega}\right>$ corresponding to the
situation where the input number of photons is kept constant when
varying the optical frequency, which was the case in the
experiment in \cite{Lodahl05-PRL}. Note that we have skipped the
indices $ab$ on the transmission coefficients in order to simplify
notation.

In the lowest order approximation, the electric field transmission
coefficient $t_{\omega}$ can be described by a circular Gaussian
process, where $T_{\omega} = \left|t_{\omega} \right|^2$. For such
Gaussian random variables, we have \cite{Goodman}
\be \overline{ t_1^* \ldots t_k^* t_{k+1} \ldots t_{2k}} =
\sum_{\pi} \overline{t_1^* t_p}  \times \overline{t_2^* t_q}
\times \ldots \times  \overline{t_k^* t_r}, \ee
where the summation is over all $k!$ permutations of the indices.
Based on this theorem, we can evaluate the moments of $T_{\omega}$
that are relevant for Eqs. (\ref{Corr-fcts-a}) and
(\ref{Corr-fcts-b}) :
\bs \bea \overline{T_{\omega} T_{\omega+\Delta \omega}} &=&
\overline{ T_{\omega}} ^2 + \left| \overline{ t_{\omega}^*
t_{\omega+\Delta
\omega}} \right|^2, \\
\overline{T_{\omega}^2 T_{\omega+\Delta \omega}} &=&
\overline{T_{\omega} T_{\omega+\Delta \omega}^2} = 2
\overline{T_{\omega}}^3 + 4 \overline{T_{\omega}} \left|
\overline{ t_{\omega}^* t_{\omega+\Delta
\omega}} \right|^2, \\
 \overline{ T_{\omega}^2 T_{\omega+\Delta \omega}^2}
&=& 4 \overline{T_{\omega}}^4 + 16 \overline{ T_{\omega}}^2 \left|
\overline{ t_{\omega}^* t_{\omega+\Delta \omega}} \right|^2 + 4
\left| \overline{ t_{\omega}^* t_{\omega+\Delta \omega}}
\right|^4, \nn \\
&& \\ \overline{ T_{\omega}^n} &=& n! \overline{ T_{\omega}}^n
 \eea \es
 First, for the quantum noise, let us restrict to
the special case of shot noise (SN), which corresponds to $F_a =
1$. Hence, the correlation functions in Eqs. (\ref{Corr-fcts-a})
and (\ref{Corr-fcts-b}) are given by
\bs \bea C_{ab}^{SN}(\Delta \omega) &=& \frac{\left| \overline{
t_{\omega}^* t_{\omega+\Delta
\omega} } \right|^2 }{{\overline{ T_{\omega} } }^2} \equiv f(\Delta \omega), \\
 C_{ab}^{CN}(\Delta \omega) &=& \frac{ \left| \overline{
t_{\omega}^* t_{\omega+\Delta \omega}} \right|^4 + 4 \overline{
T_{\omega}}^2 \left| \overline{ t_{\omega}^* t_{\omega+\Delta
\omega}} \right|^2  }{\overline{ T_{\omega}}^4} \nn \\
&\equiv& f^2(\Delta \omega) + 4 f(\Delta \omega),
 \eea \es
where
\be f(\Delta \omega) =  \frac{\Delta \omega/\omega_D}{  \cosh
(\sqrt{\Delta \omega/\omega_D}) -  \cos(\sqrt{\Delta
\omega/\omega_D}) }, \ee
is the function that describes the frequency decay of the
intensity correlations \cite{Berkovits94} with $\omega_D= D/2
L^2$.

\begin{figure}[t]
\includegraphics[width=\columnwidth]{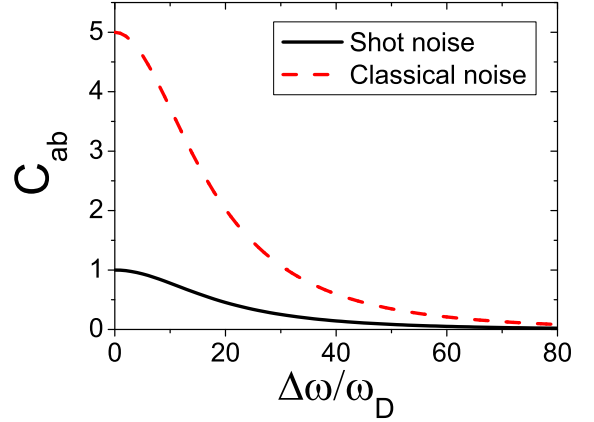}
 \caption{Noise correlation function for shot noise (full curve) and classical noise (dashed curve) as a function
 of frequency offset.  }
 \label{C-ab-SN-TN}
\end{figure}

The correlation functions for shot noise and technical noise are
plotted in Fig. \ref{C-ab-SN-TN} as a function of frequency offset
$\Delta \omega$. We observe that the noise correlations are much
stronger in the case of classical noise relative to shot noise,
thus the correlation at $\Delta \omega =0$ is a factor $5$ higher.
This pronounced difference was verified experimentally in
\cite{Lodahl05-PRL}. We note that the correlation function for
shot noise is identical to what one would get if recording the
intensity, the latter being independent of the actual quantum
state. Notably, the quantum fluctuations provide an independent
measure of frequency speckle correlations, and as will be seen in
the following, the noise correlations depend strongly on the
quantum state of light used in the experiment.

In the general case of an arbitrary quantum state characterized by
the Fano factor $F_a$, the correlation function is given by Eq.
(\ref{Corr-fcts-a}). This expression contains terms of various
orders in $\overline{T_{\omega}}$, where in general
$\overline{T_{\omega}} \ll 1,$ i.e. Eq. (\ref{Corr-fcts-a}) can be
expanded in orders of the intensity transmission coefficient. In
this expansion of the products of transmission coefficients also
higher-order contributions must be included that originate from
deviations from Gaussian statistics of the transmission
coefficients. The lowest order contribution of such mesoscopic
correlations provides a correction term in the product of two
transmission coefficients \cite{deBoer02,deBoer_thesis}
\be
 \overline{T_{\omega} T_{\omega+\Delta \omega}} =
 \overline{ T_{\omega}} ^2 + \left| \overline{ t_{\omega}^*
t_{\omega+\Delta \omega}} \right|^2 + \frac{3 L^2}{2 \ell^2}
g(\Delta \omega)\overline{T_{\omega}}^3 , \ee
where $L$ is the thickness of the random medium, $\ell$ is the
transport mean free path, and we have defined the function
\be g(\Delta \omega) = \frac{\omega_D}{\Delta \omega} \times
\frac{ \sinh (\sqrt{\Delta \omega/\omega_D}) - \sin(\sqrt{\Delta
\omega/\omega_D}) }{ \cosh (\sqrt{\Delta \omega/\omega_D}) -
\cos(\sqrt{\Delta \omega/\omega_D}) }.\ee
Expanding Eq. (\ref{Corr-fcts-a}) to first order leads to
\be  C_{ab}^{QN}(\Delta \omega) \approx C_{ab}^I(\Delta \omega) +
C_{ab}^{II}(\Delta \omega), \ee
where
\bs \bea
 C_{ab}^I(\Delta \omega) &=& f(\Delta \omega), \\
  C_{ab}^{II}(\Delta \omega)&=&  \frac{3 L^2}{2 \ell^2} g(\Delta \omega) \overline{T_{\omega}} + 4
(F_a-1) f(\Delta \omega) \overline{T_{\omega}}. \nn
\\
&& \label{Cab-2order} \eea \es
The dominating term in the expansion $(C_{ab}^I)$ is simply equal
to the contribution obtained with shot noise. Deviations from shot
noise behavior is observed when using quantum states of light
different from the coherent state, i.e. $F_a \neq 1,$ which gives
rise to the second term in Eq. (\ref{Cab-2order}). The quantum
correction competes with classical mesoscopic correlations (first
term in Eq. (\ref{Cab-2order})), which is in contrast to the
dominating quantum corrections found in the fluctuations of the
total transmission and reflection \cite{Lodahl05-corr}.

\begin{figure}[t]
\includegraphics[width=\columnwidth]{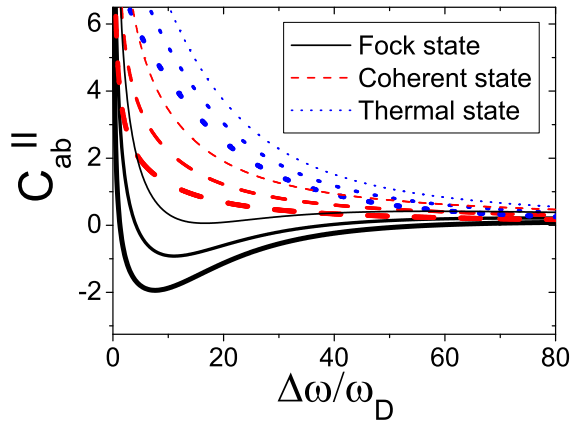}
 \caption{Second-order correlation function $C_{ab}^{II}$ normalized to $\overline{T_{\omega}}$ as
 a function of frequency offset for three different quantum states of light: Fock state $F_a=0$ (solid line),
 coherent state $F_a=1$ (dashed line) and thermal state $F_a=2$ (dotted line). The ratios of mean free path to the
 sample thickness are $\ell/L = 1/3$ (bold lines), $\ell/L = 1/4$ (medium
 lines), and $\ell/L = 1/5$ (thin lines).
 }
 \label{C-ab-II}
\end{figure}

 In Fig.
\ref{C-ab-II}, we plot the second-order noise correlation function
$C_{ab}^{II}(\Delta \omega)$ for three different quantum states of
light, corresponding to $F_a=0$, $F_a=1$, and $F_a=2.$ These Fano
factors can be achieved with single-mode Fock states, coherent
states and thermal states, respectively \cite{Lodahl05-corr}.
Figure \ref{C-ab-II} also indicates the correlation function for
different ratios $\ell/L$, where $\ell/L \ll 1$ for optical media
where multiple scattering dominates. The classical mesoscopic
correlations, which also can be extracted from intensity
measurements, are obtained for $F_a=1.$ In the limit $\Delta
\omega/\omega_D \rightarrow 0,$ the classical correlation function
diverges, which is a consequence of the plane-wave approximation
and is suppressed when including finite-width beams
\cite{deBoer02}. This sensitivity to the width of the beam only
plays an important role for the correlations at $\Delta
\omega/\omega_D \sim 1.$ We observe from Fig. \ref{C-ab-II} that
either positive or negative quantum noise correlations are
obtained using either super-Poissonian ($F_a>1$) or sub-Poissonian
photons ($F_a<1$), respectively. Consequently, the fluctuations
are found to possess novel correlations depending on the quantum
state of light, which is markedly different from intensity
correlations that are independent of the quantum state.

A novel noise correlation function for multiple scattered light
was introduced and evaluated for both classical noise and for
arbitrary single-mode quantum states. Pronounced different
correlations were found when comparing classical noise to quantum
noise. Including higher-order correction terms in an expansion in
the transmission coefficient, quantum corrections to the noise
correlation function were predicted that have no analogy in
classical intensity measurements.

I would like to thank Ad Lagendijk for fruitful discussions and
encouragement. The work is supported by the Danish Research
Agency.


\begin{thebibliography}{}

\bibitem{Feng88}
S. Feng, C. Kane, P.A. Lee, and A.D. Stone, Phys. Rev. Lett. {\bf
61}, 834 (1988).

\bibitem{Freund88}
I. Freund, M. Rosenbluh, and S. Feng, Phys. Rev. Lett. {\bf 61},
2328 (1988).

\bibitem{Garcia91}
N. Garzia and A.Z. Genack, Opt. Lett. {\bf 16}, 1132 (1991).

\bibitem{Vellekoop05}
I. Vellekoop, P. Lodahl, and A. Lagendijk, Phys. Rev. E {\bf 71},
056604  (2005).

\bibitem{Scheffold98}
F. Scheffold and G. Maret, Phys. Rev. Lett. {\bf 81}, 5800 (1998).

\bibitem{Chabanov00}
A.A. Chabanov, M. Stoytchev, and A.Z. Genack, Nature {\bf 404},
850 (2000).

\bibitem{Lodahl05-PRL}
P. Lodahl and A. Lagendijk, Phys. Rev. Lett. {\bf 94}, 153905
(2005).

\bibitem{Lodahl05-corr}
P. Lodahl, A.P. Mosk, and A. Lagendijk,
http://arxiv.org/abs/quant-ph/0502033 (2005).

\bibitem{Beenakker98} C.W.J. Beenakker, Phys. Rev. Lett.
{\bf 81}, 1829 (1998).

\bibitem{Goodman}
J.W. Goodman, \textit{Statistical Optics} (John Wiley \& Sons, New
York, 1985).

\bibitem{Berkovits94}
R. Berkovits and S. Feng, Phys. Rep. {\bf 238}, 135 (1994).

\bibitem{deBoer02}
J.F. de Boer, M.P. van Albada, and A. Lagendijk, Phys. Rev. B {\bf
45}, 658 (1992).

\bibitem{deBoer_thesis}
J.F. de Boer, \textit{Optical fluctuations on the transmission and
reflection of mesoscopic systems}, (Ph.D. Thesis, University of
Amsterdam, 1995), available on:
www.tn.utwente.nl/cops/pdf/theses/deboer.pdf.


\end{thebibliography}
\end{document}